\documentstyle[aps,preprint]{revtex}

\newcommand{\Ord}[1]{{\cal O}\left(#1\right)}

\newcommand{\tr}{{\, \rm tr}}

\tightenlines

\begin{document}

\title{Integration over matrix spaces with unique invariant measures}
\author{T. Prosen$^{1,3}$, T. H. Seligman$^{2,3}$ and
H. A. Weidenm\"uller$^3$}

\address{$^1$ Physics Department, Faculty of Mathematics and Physics, 
University of Ljubljana, Slovenia\\
$^2$ Centro de Ciencias F\'\i sicas, University of Mexico (U.N.A.M.),
Cuernavaca, and Centro Internacional de Ciencias, Cuernavaca, Mexico\\
$^3$ Max--Planck--Institut f\" ur Kernphysik, Heidelberg, Germany
}

\maketitle

\begin{abstract}
We present a method to calculate integrals over monomials of matrix
elements with invariant measures in terms of Wick contractions. The
method gives exact results for monomials of low order. For
higher--order monomials, it leads to an error of order $1/N^{\alpha}$
where $N$ is the dimension of the matrix and where $\alpha$ is
independent of the degree of the monomial. We give a lower bound on
the integer $\alpha$ and show how $\alpha$ can be increased
systematically. The method is particularly suited for symbolic
computer calculation. Explicit results are given for ${\rm O}(N), {\rm
  U}(N)$ and for the circular orthogonal ensemble.
\end{abstract}

\section{Introduction}
\label{int}

The calculation of group integrals of monomials of matrix elements
for compact Lie groups has a long tradition going back to Ullah and
Porter~\cite{ullah}. Their results were later extended~\cite{MS} to
the space of symmetric unitary matrices endowed with Dyson's invariant
measure~\cite{dyson}. The problem was nearly dormant for some years
but was recently solved completely for the orthogonal group by
recursion~\cite{gorin}.  Unfortunately, it seems that there is no
easy generalization of this method to other groups. The method
proposed in Ref.~\cite{MS}, on the other hand, is quite general but
soon becomes rather cumbersome. Moreover, that method is not suited
for computer--supported analytical work.

Aside from their immanent group--theoretical significance, group
integrals over monomials of matrix elements for compact Lie groups
are important in applications of random--matrix theory. This field has
seen an explosive growth over the last decade and has become an
important tool in a great variety of fields in physics, chemistry and
related areas~\cite{NT}. This fact lends further urgency to the
evaluation of the above--mentioned group integrals.

In this paper, we present and analyse a novel approach to the problem.
We encountered the problem in the study of a random--matrix model for
a class of chaotic systems (the semi--separable systems)~\cite{PSW}.
We found that we could use the invariants of the orthogonal group to
construct a weight function.  With the help of this function, it was
possible to evaluate the group integrals in question by simple Wick
contractions. We implemented the scheme on the computer and found that
beyond the expected exact results for low--order monomials, monomials
of higher order were also calculated correctly up to and including the
sub--leading order in $1/N$ where $N$ is the dimension of the matrices
under consideration. We conjectured that this statement holds for
monomials of any order. It is the purpose of the present paper to
extend and prove the conjecture and to explore the scope of its
validity beyond the orthogonal group.

The group integrals extend over a compact matrix space with a measure
which is uniquely determined by the underlying symmetry group. In
order to reduce the computation to Wick contractions, we consider an
extended matrix space where all matrix elements are independent
Gaussian variables. In this space, all integrations trivially reduce
to Wick contractions and can easily be implemented in many programming
languages. The constraints due to the group structure are then
introduced in an approximate fashion through a weight function $w$
appearing as factor in the integrand. This function is chosen in such
a way that the integrals yield the exact values for the lowest--order
invariants of the group. It turns out that $w$ is not always positive
and, thus, not a measure. This, however, is not a significant obstacle.

Candidates for our spaces are the orthogonal group ${\rm O}(N)$, the
unitary group ${\rm U}(N) = {\rm CUE}(N)$, also known as the circular
unitary ensemble, the circular orthogonal ensemble of symmetric
unitary matrices ${\rm COE}(N)$, and the circular symplectic ensemble
which is isomorphic to the unitary symplectic group ${\rm CSE}(N) =
{\rm USP}(2N)$. We note that ${\rm COE}(N)$ is not a group --- for
all other cases we can use the Haar measure, while in this case we
have to use Dyson's invariant measure. 

We first present our arguments for the case of the orthogonal group.
In Section~\ref{wei} we construct the weight function $w$ from the
invariants of the orthogonal group. We show that the defining
equations for $w$ always have a unique solution. We give explicit
expressions for $w$ in the simplest cases. We show that low--order
monomials are calculated exactly using Wick contraction. In
Section~\ref{exp}, we show that monomials of higher order are
evaluated correctly by Wick contraction, up to an error of order
$N^{- \alpha}$. We establish a lower bound for the exponent $\alpha$.
In Section~\ref{uni} we extend our arguments to other matrix spaces.
We give explicit expressions for the weight functions for ${\rm U}(N)$
and for ${\rm COE}(N)$. The more involved and less important case
of CSE is only touched upon.

\section{The Weight Function for ${\rm O}(N)$}
\label{wei}

As explained in Section~\ref{int}, we start with a space of real
matrices $M$. The elements are taken as independent
Gaussian--distributed variables with zero mean and identical
variances. In other words, our measure $d\mu_g$ for integration is the
product of the differentials of all matrix elements times ${\cal N}
\exp \{- N \ {\rm trace} (M M^T)\}$ where ${\cal N}$ is a normalization
factor. We recall that $N$ is the dimension of the matrices $M$. We
are interested in values of $N \gg 1$. The measure is invariant under
right or left multiplication of $M$ with any orthogonal matrix. To
restrict the integration to the orthogonal group, we could think of
multiplying $d\mu_g$ with a set of delta functions expressing all the
constraints due to orthonormality. This is clearly impractical.
Instead, we modify the measure by multiplying $d\mu_g$ with a weight
function $w_{\kappa}$. This function is chosen in such a way that all
orthogonal invariants up to and including order $2 \kappa$ are exactly
reproduced when we use $w_{\kappa}$ as a weight function under the
integrals over $M$. We note in passing that our present notation
differs from that of Ref.~\cite{PSW}. Our index $\kappa$ equals two
times the index $k$ used there.

To determine $w_{\kappa}$, we consider all invariants $I_j(O)$ up to
order $2 \kappa$ in the matrix elements $O$ of the orthogonal group.
Here, $j$ is a running index. We recall that all such invariants are
even in the $O$'s. In every invariant $I_j(O)$, we replace $O$ by $M$
to obtain $I_j(M)$. We write $w_{\kappa}$ as a linear combination of
all the $I_j(M)$'s up to order $2 \kappa$ in $M$. The coefficients of
the linear combination are determined by the requirement that the
average of every $I_j(M)$, calculated by integration over $d\mu_g$
with weight function $w_{\kappa}$, yields the same result as
integration of $I_j(O)$ over the orthogonal group. We recall that
with $n$ a positive integer, the invariants of the orthogonal group
are given by expressions of the form $\tr \{(O O^T)^n \}$, or by
products of such expressions. We accordingly write the quantities
$I_j(M)$ in the form
\begin{equation}
I^{(k)}_{\bf k}(M) = \prod_{k_i} \tr \{(MM^T)^{k_i}\} \ .
\label{eq:inv}
\end{equation}
Here $2 k$ denotes the degree of $I$ in $M$, and ${\bf k} = (k_1, k_2,
\ldots)$ is a partition of $k$ into positive integers $k_i \ge 1$ with
$k_1 + k_2 + \ldots = k$. Without loss of generality we require that
$k_1 \geq k_2 \geq \ldots$. The weight function $w_{\kappa}$ is now
written as
\begin{equation}
w_{\kappa}(M) = a^{(\kappa)}_0 + \sum_{k=1}^{\kappa} \sum_{\bf k}
a^{(\kappa)}_{\bf k} I^{(k)}_{\bf k}(M) \ .
\label{eq:weight}
\end{equation}
The sum on the right--hand side of Eq.~(\ref{eq:weight}) extends over a
complete set of linearly independent invariants up to order $2 \kappa$
in $M$.

We determine the coefficients $a^{(\kappa)}_{\bf k}$ from the
conditions of orthonormality. More precisely, we require that the
relations
\begin{eqnarray}
\int d\mu_g w_{\kappa}(M) &=& 1, \nonumber \\
\int d\mu_g w_{\kappa}(M) (M M^T)_{i_1j_1} &=& \delta_{i_1,j_1}
\ , \nonumber \\
\int d\mu_g w_{\kappa}(M) (M M^T)_{i_1j_1} (M M^T)_{i_2j_2} &=&
\delta_{i_1,j_1}\delta_{i_2,j_2} \ , \nonumber \\
\ldots \nonumber \\
\int d\mu_g w_{\kappa}(M) (M M^T)_{i_1j_1} (M M^T)_{i_2j_2} \ldots
(M M^T)_{i_{\kappa}j_{\kappa}} &=&
\delta_{i_1,j_1} \delta_{i_2,j_2} \times \ldots \times
\delta_{i_{\kappa},j_{\kappa}}
\label{eq:cond}
\end{eqnarray}
be fulfilled identically. Relations of the form~(\ref{eq:cond}) hold
for any value of $\kappa$ for the orthogonal group but must be imposed
for the integration over the matrices $M$.

Eqs.~(\ref{eq:cond}) determine the coefficients $a^{(\kappa)}_{\bf k}$
uniquely. To show this, we take traces over these equations in such a
way that the integrals on the left--hand sides take the form $\int
d\mu_g w_{\kappa}(M) I^{(k)}_{\bf k}(M)$. The resulting set of
equations has the form
\begin{equation}
\int d\mu_g w_{\kappa}(M) I^{(k)}_{\bf k}(M) = B^{(k)}_{\bf k}
\label{eq:cond1}
\end{equation}
where the coefficients $B^{(k)}_{\bf k}$ are given by powers of $N$,
with $N$ the dimension of the matrices $M$. Recalling
Eq.~(\ref{eq:weight}), we see that Eqs.~(\ref{eq:cond1}) constitute a
set of linear equations for the coefficients $a^{(\kappa)}_{\bf k}$.
There are obviously as many equations as there are coefficients
$a^{(\kappa)}_{\bf k}$. We conclude that Eqs.~(\ref{eq:cond1}) possess
a unique solution unless the determinant of the matrix $C$ with elements
$C^{(k_1 k_2)}_{{\bf k_1} {\bf k_2}} = \int d\mu_g I^{(k_1)}_{\bf k_1}
I^{(k_2)}_{\bf k_2}$ vanishes. But if $\det ( C^{(k_1 k_2)}_{{\bf k_1}
{\bf k_2}} ) = 0$, there exists a nontrivial solution
$b^{(k_2)}_{\bf k_2}$ of the homogeneous equation $\sum_{k_2 {\bf k}_2}
C^{(k_1 k_2)}_{{\bf k_1} {\bf k_2}} b^{(k_2)}_{\bf k_2} = 0$.
The existence of this solution implies that we also have $\sum_{k_1
  k_2} \sum_{{\bf k}_1 {\bf k}_2} b^{(k_1)}_{\bf k_1} C^{(k_1
  k_2)}_{{\bf k_1} {\bf k_2}} b^{(k_2)}_{\bf k_2} = 0$. Recalling the
definition of the matrix $C$, we observe that the last relation can be
written as $\int d\mu_g |\sum_{k_1 {\bf k}_1} I^{(k_1)}_{\bf k_1}
b^{(k_1)}_{\bf k_1}|^2 = 0$. But the integrand in the last expression
is positive semidefinite and does not vanish identically. Therefore,
it is not possible that $\det ( C^{(k_1 k_2)}_{{\bf k_1} {\bf k_2}} )$
vanishes, and the solution of Eqs.~(\ref{eq:cond1}) exists and is
unique. This solution also solves Eqs.~(\ref{eq:cond}). To see this,
let us assume the contrary and focus attention on the second of
Eqs.~(\ref{eq:cond}). (The argument is easily extended to the entire
set of Eqs.~(\ref{eq:cond})). Inserting the solution of
Eqs.~(\ref{eq:cond1}) into the left--hand side of that equation yields
on the right--hand side the terms $\delta_{i_1 j_1} + A_{i_1j_1}$
where the matrix $A$ is both traceless and invariant under every
orthogonal transformation. This implies $A = 0$, in contradiction to
the assumption that we did not find a solution of the second of
Eqs.~(\ref{eq:cond}).

Eqs.~(\ref{eq:cond}) imply that the integrals over all polynomials of
degree $n \leq 2 \kappa$ in $M$ are equal to the corresponding
expressions for ${\rm O}(N)$. To see this, it suffices to consider the
integral over an arbitrary monomial of degree $n$. It is obvious that
the integral vanishes unless $n$ is even, $n = 2 k$. We write the
monomial as ${\cal M}^{(n)} = M_{i_1 j_1} M_{i_2 j_2} \ldots M_{i_n
j_n}$. The integral over ${\cal M}^{(n)}$ is obviously invariant under
right or left multiplication with any orthogonal transformation.
Therefore, the integral over ${\cal M}^{(n)}$ must be a linear
combination of invariants multiplied by a suitable set of Kronecker
deltas in the indices $i_1, \ldots, i_n$ and $j_1, \ldots, j_n$. By
construction the invariants have the same values as in ${\rm O}(N)$.

Inspection shows that the weight function $w_0 = 1$ fulfills the
second of Eqs.~(\ref{eq:cond}) automatically. Thus, $w_1 = w_0$ and,
therefore, $a^{(1)}_0 = 1, \ a^{(1)}_1 = 0$. The first nontrivial
condition is, therefore, the one appearing in line 3 of
Eqs.~(\ref{eq:cond}). This condition (and all that follow below it) is
violated by $w_0$. We now give the explicit results for the first few
weight functions $w_{\kappa}$. These were obtained with the help of
the Mathematica program. For $\kappa = 2$, we find
\begin{eqnarray}
a^{(2)}_0 &=& 1 - \frac{N^2}{4} \nonumber \\
a^{(2)}_1 &=& \frac{N}{2} \nonumber \\
a^{(2)}_2 &=& -\frac{N^3}{4(-1 + N)(2 + N)} \nonumber \\
a^{(2)}_{1 1} &=& \frac{N^2}{4(-1 + N)(2 + N)} \ .
\label{eq1a}
\end{eqnarray}
For $\kappa = 3$, we have
\begin{eqnarray}
a^{(3)}_0 &=& 1 - \frac{7 N^2}{12} \nonumber \\
a^{(3)}_1 &=& \frac{3N}{2} \nonumber \\
a^{(3)}_2 &=& -\frac{5 N^3}{4(-1 + N)(2 + N)} \nonumber \\
a^{(3)}_{1 1} &=& \frac{5 N^2}{4(-1 + N)(2 + N)} \nonumber \\
a^{(3)}_3 &=& \frac{N^5}{3(-2 + N)(-1 + N)(2 + N)(4 + N)} \nonumber \\
a^{(3)}_{2 1} &=&-\frac{N^4}{(-2 + N)(-1 + N)(2 + N)(4 + N)} \nonumber
\\ 
a^{(3)}_{1 1 1} &=& \frac{2 N^3}{3(-2 + N)(-1 + N)(2 + N)(4 + N)} \ .
\label{eq2a}
\end{eqnarray}
For $\kappa = 4$, we have
\begin{eqnarray}
a^{(4)}_0 &=& 1 - \frac{23\,N^2}{24} + \frac{N^4}{32} \nonumber \\
a^{(4)}_1 &=& 3\,N - \frac{N^3}{8} \nonumber \\
a^{(4)}_2 &=& \frac{-60\,N^3 + N^5}{16\,\left( -1 + N \right) \,
  \left( 2 + N \right) } \nonumber \\
a^{(4)}_{1 1} &=& \frac{56\,N^2 + 2\,N^3 + N^4}
  {16\,\left( -1 + N \right) \,\left( 2 + N \right) } \nonumber \\
a^{(4)}_3 &=& \frac{7\,N^5}{3\,\left( -2 + N \right) \,\left( -1 + N
  \right) \, \left( 2 + N \right) \,\left( 4 + N \right) } \nonumber
  \\
a^{(4)}_{2 1} &=& \frac{-48\,N^4 - 2\,N^5 - N^6}
  {8\,\left( -2 + N \right) \,\left( -1 + N \right) \,
    \left( 2 + N \right) \,\left( 4 + N \right) } \nonumber \\
a^{(4)}_{1 1 1} &=& \frac{88\,N^3 + 6\,N^4 + 3\,N^5}
  {24\,\left( -2 + N \right) \,\left( -1 + N \right) \,
    \left( 2 + N \right) \,\left( 4 + N \right) } \nonumber \\
a^{(4)}_4 &=& -\frac{ N^7\,\left( 6 + 5\,N \right) }
  {8\,\left( -3 + N \right) \,\left( -2 + N \right) \,
    \left( -1 + N \right) \,\left( 1 + N \right) \,\left( 2 + N
    \right) \, \left( 4 + N \right) \,\left( 6 + N \right) } \nonumber
  \\
a^{(4)}_{3 1} &=& \frac{N^6\,\left( 6 + 5\,N \right) }
  {2\,\left( -3 + N \right) \,\left( -2 + N \right) \,
    \left( -1 + N \right) \,\left( 1 + N \right) \,\left( 2 + N
  \right) \, \left( 4 + N \right) \,\left( 6 + N \right) } \nonumber
  \\
a^{(4)}_{2 2} &=& \frac{N^7\,\left( 18 + 5\,N + N^2 \right) }
  {32\,\left( -3 + N \right) \,\left( -2 + N \right) \,
    \left( -1 + N \right) \,\left( 1 + N \right) \,\left( 2 + N
  \right) \, \left( 4 + N \right) \,\left( 6 + N \right) } \nonumber
  \\
a^{(4)}_{2 1 1} &=& -\frac{N^5\,\left( 72 + 78\,N + 5\,N^2 + N^3
  \right) } {16\,\left( -3 + N \right) \,\left( -2 + N \right) \,
    \left( -1 + N \right) \,\left( 1 + N \right) \,\left( 2 + N
  \right) \, \left( 4 + N \right) \,\left( 6 + N \right) } \nonumber
  \\
a^{(4)}_{1 1 1 1} &=& \frac{N^4\,\left( 72 + 78\,N + 5\,N^2 + N^3
  \right) } {32\,\left( -3 + N \right) \,\left( -2 + N \right) \,
    \left( -1 + N \right) \,\left( 1 + N \right) \,\left( 2 + N
  \right) \,  \left( 4 + N \right) \,\left( 6 + N \right) \ . }
\label{eq3a}
\end{eqnarray}
We note that with increasing $\kappa$, the expressions become rather
involved. Moreover, the coefficients $a^{(\kappa)}_{\bf k}$ with the
same lower indices ${\bf k}$ change with $\kappa$.

\section{Monomials of Higher Order for ${\rm O}(N)$: The $1/N$
  Expansion}
\label{exp}

We have seen that integrals over all polynomials of degree $n \leq
2 \kappa$ have the same values as for ${\rm O}(N)$. What about
polynomials of higher order? Again, it suffices to consider monomials
${\cal M}^{(k)}$ of even degree $2 k$ with $k > \kappa$. We show that 
the integral over ${\cal M}^{(k)}$ coincides with the result for
${\rm O}(N)$ up to terms of order $N^{- \alpha}$ where the integer
exponent $\alpha$ is positive and independent of $k$. More precisely,
we show that for $k > {\kappa}$, we have
\begin{equation}
N^{k}\int d\mu_g w_{\kappa}(M) \prod_{\nu=1}^{2k} M_{i_\nu j_\nu} =
N^{k}\int dh_{{\rm O}(N)} \prod_{\nu=1}^{2k} O_{i_\nu j_\nu} +
\Ord{1/N^{\alpha}} \ {\rm where} \ \alpha \geq
[{\kappa}/2]+1 \ .
\label{eq:result}
\end{equation}
Here $[{\kappa}/2]$ indicates the integer part of ${\kappa}/2$, and
$dh_{{\rm O}(N)}$ denotes the Haar measure for integration over ${\rm
  O}(N)$. We note that the factors $N^{k}$ in front of the integrals
normalise the $N$-dependence so that these terms are (at most) of
order $1$. Another equivalent form of Eq.~(\ref{eq:result}) is
obtained by summing over pairs of indices $j_1 = j_2, j_3 = j_4,
\ldots$. This removes the factors $N^k$ and yields
\begin{equation}
\int d\mu_g w_{\kappa}(M) \prod_{\nu=1}^{k} (M M^T)_{i_{\nu} l_{\nu}}
= \int dh_{{\rm O}(N)} \prod_{\nu=1}^{k} (O O^T)_{i_\nu l_\nu} +
\Ord{1/N^{\alpha}} \ {\rm where} \ \alpha \geq
[{\kappa}/2]+1 \ .
\label{eq:result1}
\end{equation}
The equivalence of Eq.~(\ref{eq:result}) and Eq.~(\ref{eq:result1}) 
follows from the fact the matrix $C$ discussed above, if defined with 
respect to properly scaled monomials, does not depend on $N$. The
remainder of this Section is devoted to proving Eq.~(\ref{eq:result1}).

It is useful to introduce a few auxiliary concepts. We consider
Gaussian integrals over monomials of $M$ without the weight function
$w_{\kappa}$. We write for brevity
\begin{equation}
\int d\mu_g \prod_{\nu=1}^{k} (M M^T)_{i_{\nu} l_{\nu}} = \langle
(M M^T)^{k} \rangle_g
\label{eq:defg}
\end{equation}
where the index $g$ indicates the purely Gaussian integration. To
define the completely correlated part of this expression, we consider
first the case $k = 2$. We use Wick contraction and have
\begin{equation}
\langle (M M^T)^{2} \rangle_g = \langle (M M^T) \rangle_g \langle
(M M^T) \rangle_g + \langle (M M^T)^{2} \rangle_{gc} \ .
\label{eq:defgc}
\end{equation}
The last term on the right--hand side of Eq.~(\ref{eq:defgc}) is the
completely correlated term. For the general case of arbitrary order $2
k$, we define the correlated part $\langle (M M^T)^{k} \rangle_{gc}$
as that contribution to $\langle (M M^T)^{k} \rangle_{g}$ which cannot
be written in the form of products of two or more factors, each of which is a
complete Wick contraction of powers of $M M^T$. It is easy to see that
\begin{equation}
\langle (M M^T)^{k} \rangle_{gc} = \Ord{1/N^{k-1}} \ .
\label{eq:ord}
\end{equation}
The linear increase with $k$ in inverse powers of $N$ in
Eq.~(\ref{eq:ord}) is due to the fact that every Wick contraction
which connects two $M$'s appearing in different factors $M M^T$
suppresses one summation index. Therefore, the correlated part
$\langle (M M^T)^{k} \rangle_{gc}$ contributes the highest--order
terms in $1/N$ to $\langle (M M^T)^{k} \rangle_{g}$.

We now consider integrals involving the weight function $w_{\kappa}$
and use the same notation,
\begin{equation}
\int d\mu_g w_{\kappa} \prod_{\nu=1}^{k} (M M^T)_{i_{\nu} l_{\nu}} =
\langle w_{\kappa} (M M^T)^{k} \rangle_g \ .
\label{eq:def1g}
\end{equation}
Again using Wick contraction, we define the correlated part $\langle
w_{\kappa} (M M^T)^{k} \rangle_{gc}$ of this expression as that part
which cannot be written as the product of two or more factors, each of
which is a complete Wick contraction of powers of $M M^T$ and/or
$w_{\kappa}$.

We proceed to show that in the equations relating the integral
$\langle w_{\kappa} (M M^T)^{k} \rangle_g$ to the integral over the
Haar measure, the leading correction term (lowest order in $1/N$)
which does not cancel is given by
\begin{equation}
\langle w_\kappa (M M^T)^k \rangle_{gc}  = \Ord{1/N^{[(k+1)/2]}}
\qquad {\rm for} \qquad 1 < k \leq \kappa \ .
\label{eq:ord1}
\end{equation}
This relation is based upon the assumption that there is no accidental
cancellation among the terms contributing to lowest order in $1/N$.
Therefore, $[(k+1)]/2$ actually constitutes a lower bound on the
exponent of $1/N$.

To prove the relation~(\ref{eq:ord1}), we rewrite the defining
equations for $w_{\kappa}$, Eqs.~(\ref{eq:cond}), as follows. We
consider the expression $\langle w_{\kappa} (M M^T)^k \rangle_g$ with
$k$ integer and $k \leq \kappa$. We decompose this expression into
correlated contributions. These originate from all partitions ${\bf k}
= (k_1, k_2, \ldots)$ of $k$ with $k_i \geq k_{i+1}$ and $\sum_i k_i =
k$. We denote by $i_0$ the smallest index for which all $k_i$ with $i
> i_0$ are equal to one. Then, we have
\begin{equation}
\langle w_{\kappa} (M M^T)^k \rangle_g = \sum_{\bf k} \prod_i \langle
(M M^T)^{k_i} \rangle_{gc} + \sum_{{\bf k} \neq {\bf 1}^k}
\sum_i^{i_0} \langle w_{\kappa} (M M^T)^{k_i} \rangle_{gc} \prod_{j
  \neq i} \langle (M M^T)^{k_j} \rangle_{gc} \ .
\label{eq:dec}
\end{equation}
In the first term on the right--hand side, we have used that $\langle
w_{\kappa} \rangle_g = 1$. In the second term, we have used that
$\langle w_{\kappa} (M M^T) \rangle_{gc} = 0$. Trivially, the second
sum on the right--hand side of Eq.~(\ref{eq:dec}) does not extend over
the partition ${\bf 1}^k = (1,1,1,\ldots)$ ($k$ terms unity).
According to Eqs.~(\ref{eq:cond}), the expression in
Eq.~(\ref{eq:dec}) equals $(\langle (M M^T) \rangle_g )^k$. This
equals the contribution from the first sum on the right--hand side for
the partition ${\bf 1}^k$. The remaining terms must vanish,
\begin{equation}
\sum_{{\bf k} \neq {\bf 1}^k} \prod_i \langle (M M^T)^{k_{i}}
\rangle_{gc} + \sum_{{\bf k} \neq {\bf 1}^k} \sum_i^{i_0} \langle
w_{\kappa} (M M^T)^{k_i} \rangle_{gc} \prod_{j \neq i} \langle (M
M^T)^{k_{j}} \rangle_{gc} = 0 \ .
\label{eq:dec1}
\end{equation}
Eq.~(\ref{eq:dec1}) must hold for all values of $k$ with $k \leq
\kappa$. To proceed, we observe that the partitions of $k$ can be
grouped into classes as follows: Partitions within the same class carry the same
number $p$ of $k_i$'s that have value unity. The classes are labeled
by $p$, namely ${\cal C}_p$. 
For instance, for $k = 6$, class ${\cal C}_2$ contains the
partitions $(4,1,1)$ and $(2,2,1,1)$. In Eq.~(\ref{eq:dec1}), we order
the sum over ${\bf k}$ by grouping together all partitions which
belong to the same class. We show presently that each such
contribution must vanish separately. Then, we have for every $p = 
0,1,\ldots,k-2$ that
\begin{equation}
\sum_{{\bf k} \neq {\bf 1}, {\bf k} \in {\cal C}_p} \prod \langle (M M^T)^{k_i}
\rangle_{gc} + \sum_{{\bf k} \in {\cal C}_p} \sum_i^{i_0} \langle w_{\kappa}
(M M^T)^{k_i} \rangle_{gc} \prod_{j \neq i} \langle (M M^T)^{k_j}
\rangle_{gc} = 0 \ .
\label{eq20}
\end{equation}
Eq.~(\ref{eq20}) follows directly from the facts that
Eq.~(\ref{eq:dec1}) holds for all $k \leq \kappa$, and that the
contributions from class ${\cal C}_p$ to a partition of $k$ are the same as the
contributions of class ${\cal C}_0$ to a partition of $k - p$, except for a
string of Kronecker delta's due to the factors $(\langle (M M^T)
\rangle_g)^p$.

We are now in the position to prove the relation~(\ref{eq:ord1}). We
observe that in Eqs.~(\ref{eq20}), the term $\langle w_{\kappa} (M
M^T)^k \rangle_{gc}$ appears only in the class ${\cal C}_0$. Therefore, we
have
\begin{eqnarray}
\langle w_{\kappa} (M M^T)^k \rangle_{gc} &=& - \sum_{{\bf k} \in {\cal C}_0}
\prod_i \langle (M M^T)^{k_i} \rangle_{gc}
\nonumber \\
&&- \sum_{{\bf k} \in {\cal C}_0, {\bf k} \neq (k)} \sum_i \langle w_{\kappa}
(M M^T)^{k_i} \rangle_{gc} \prod_{j \neq i} \langle (M M^T)^{k_j}
\rangle_{gc} \ .
\label{eq21}
\end{eqnarray}
Using complete induction, i.e., assuming that the
relation~(\ref{eq:ord1}) holds for all values of $k'$ with $k' < k$,
and using Eq.~(\ref{eq:ord}), we conclude from Eq.~(\ref{eq21}) that
the relation~(\ref{eq:ord1}) also holds for $k' + 1 = k$. The terms of
lowest order in $1/N$ originate from partitions which have either the
form $(2,2,2,\ldots)$ (for even $k$) or $(3,2,2,2,\ldots)$ (for odd
$k$). Again, we cannot rule out the occurrence of accidental
cancellations which would increase the power $[(k+1)]/2$ in the
relation~(\ref{eq:ord1}).

Having established the relation~(\ref{eq:ord1}), we turn to the center
piece of this Section, Eq.~(\ref{eq:result1}). We first consider the
case $k = \kappa + 1$ and decompose the integral into correlated terms
with contributions from all partitions of $\kappa + 1$. For these
partitions, we write ${\bf K + 1} = (k_1,k_2,\ldots)$ with $k_1 + k_2
+ \ldots = \kappa + 1$ and $k_1 \geq k_2 \geq \ldots$. We use the
notation introduced above.
Then,
\begin{eqnarray}
\langle w_{\kappa} (M M^T)^{\kappa + 1} \rangle_g &=& \sum_p
\sum_{({\bf K + 1}) \in {\cal C}_p } 
\prod_i \langle (M M^T)^{k_i} \rangle_{gc} 
\nonumber \\
&&+ \sum_p \sum_{({\bf K + 1}) \in {\cal C}_p } \sum_i^{i_0} \langle w_{\kappa}
(M M^T)^{k_i} \rangle_{gc} \prod_{j \neq i} \langle 
(M M^T)^{k_j} \rangle_{gc} \ .
\label{eq22}
\end{eqnarray}
Eqs.~(\ref{eq20}) imply that all terms with $p \neq 0$ and $p \neq
\kappa + 1$ vanish, and we are left with
\begin{eqnarray}
\langle w_{\kappa} (M M^T)^{\kappa + 1} \rangle_g &=& (\langle (M M^T) 
\rangle_g)^{\kappa + 1} + \sum_{({\bf K + 1}) \in {\cal C}_0 } \prod_i \langle
(M M^T)^{k_i} \rangle_{gc} 
\nonumber \\
&&+ \sum_{({\bf K + 1}) \in {\cal C}_0 } \sum_i \langle w_{\kappa} (M
M^T)^{k_i} \rangle_{gc} \prod_{j \neq i} \langle  (M
M^T)^{k_j} \rangle_{gc} \ .
\label{eq23}
\end{eqnarray}
We use the same argument as in the previous paragraph and the
result~(\ref{eq:ord1}) and Eq.~(\ref{eq:ord}). We conclude that the
contributions of lowest order in $1/N$ result from the partitions
$(2,2,2,\ldots)$ or $(3,2,2,2,\ldots)$, respectively, and arrive at
Eq.~(\ref{eq:result1}). The exponent $\alpha$ has the value $\alpha
\geq [\kappa/2] + 1$. In writing the inequality sign, we again allow
for the possibility that an accidental cancellation of contributions
from these partitions occurs.

We finally have to consider the case where $k - \kappa = n > 1$.
Decomposing the expression $\langle w_{\kappa} (M M^T)^{\kappa + n}
\rangle_g$ as in Eq.~(\ref{eq22}), we easily see that the terms of
lowest nonvanishing order in $1/N$ stem from the partitions in class
${\cal C}_{n - 1}$. But these give exactly the same contributions as those
for $k = \kappa + 1$ that were estimated in the last paragraph. This
completes the proof of Eq.~(\ref{eq:result1}).

\section{The unitary group and other unitary ensembles}
\label{uni}

In this Section, we primarily address integrals over momomials of
unitary matrices $U$ with respect to the Haar measure of the unitary
group. We start with Gaussian integrals over complex matrices $M$. The
real and imaginary parts of the matrix elements are independent and
Gaussian--distributed. The integrals are again worked out using Wick
contractions. For the weight function $w^u_{\kappa}$, we write in
analogy to Eq.~(\ref{eq:weight})
\begin{equation}
w^u_{\kappa}(M) = b^{(\kappa)}_0 + \sum_{k=1}^{\kappa} \sum_{\bf k}
b^{(\kappa)}_{\bf k} I^{(k)}_{\bf k}(M) \ .
\label{eq:weight1}
\end{equation}
The invariants are defined as in Eq.~(\ref{eq:inv}) with $M M^T$
replaced by $M M^{\dagger}$ where $^{\dagger}$ stands for Hermitean
conjugation. The upper index $u$ refers to the unitary case. The
values of the coefficients $b^{\kappa}_{\bf k}$ are, of course, not
the same as for the orthogonal case. The arguments for the existence
and uniqueness of the solutions carry over without change, and again
computer programs are available to perform the contractions and
calculate the coefficients $b^{\kappa}_{\bf k}$. For $w^u_2$ and
$w^u_4$ we find
\begin{eqnarray}
b^2_0 &=& 1 - \frac{N^2}{2} \nonumber \\
b^2_1 &=& N \nonumber \\
b^2_2 &=& -\frac{N^3}{2(N-1)(N+1)} \nonumber \\
b^2_{1 1} &=& \frac{N^2}{2(N-1)(N+1)} \ ,
\end{eqnarray}
and
\begin{eqnarray}
b^4_0 &=& \frac{24 - 46 \, N^2 + 3 \, N^4}{24} \nonumber \\
b^4_1 &=& \frac{-\left( N\,\left( -12 + N^2 \right)  \right) }{2}
\nonumber \\
b^4_2 &=& \frac{N^3 \, \left( -30 + N^2 \right) }
  {4\,\left( -1 + N \right) \,\left( 1 + N \right) } \nonumber \\
b^4_{1 1} &=& \frac{N^2 \, \left( 28 + N^2 \right) }
  {4\,\left( -1 + N \right) \,\left( 1 + N \right) } \nonumber \\
b^4_3 &=& \frac{14\,N^5}{3\,\left( -2 + N \right) \, \left( -1 + N
\right) \, \left( 1 + N \right) \,\left( 2 + N \right) } \nonumber \\
b^4_{2 1} &=& \frac{-N^4\,\left( 24 + N^2 \right)  }
  {2\,\left( -2 + N \right) \,\left( -1 + N \right) \, \left( 1 + N
  \right) \, \left( 2 + N \right) } \nonumber \\
b^4_{1 1 1} &=& \frac{N^3\,\left( 44 + 3\,N^2 \right) }
  {6\,\left( -2 + N \right) \,\left( -1 + N \right) \, \left( 1 + N
  \right) \, \left( 2 + N \right) } \nonumber \\
b^4_4 &=& \frac{-5\,N^7}{4\,\left( -3 + N \right) \, \left( -2 + N
  \right) \, \left( -1 + N \right) \,\left( 1 + N \right) \,\left( 2 +
  N \right) \, \left( 3 + N \right) } \nonumber \\
b^4_{3 1} &=& \frac{5\,N^6}{\left( -3 + N \right) \,\left( -2 + N
  \right) \, \left( -1 + N \right) \,\left( 1 + N \right) \,\left( 2 +
  N \right) \, \left( 3 + N \right) } \nonumber \\
b^4_{2 2} &=& \frac{N^6\,\left( 6 + N^2 \right) }
  {8\,\left( -3 + N \right) \,\left( -2 + N \right) \,
    \left( -1 + N \right) \,\left( 1 + N \right) \,\left( 2 + N
    \right) \, \left( 3 + N \right) } \nonumber \\
b^4_{2 1 1} &=& \frac{-N^5\,\left( 36 + N^2 \right) }
  {4\,\left( -3 + N \right) \,\left( -2 + N \right) \,
    \left( -1 + N \right) \,\left( 1 + N \right) \,\left( 2 + N
    \right) \, \left( 3 + N \right) } \nonumber \\
b^4_{1 1 1 1} &=& \frac{N^4\,\left( 36 + N^2 \right) }
  {8\,\left( -3 + N \right) \,\left( -2 + N \right) \,
    \left( -1 + N \right) \,\left( 1 + N \right) \,\left( 2 + N
    \right) \, \left( 3 + N \right) } \ .
\end{eqnarray}

The arguments determining the leading contribution to the $1/N$
expansion are the same ones as for the orthogonal group. Thus, we have
to increase ${\kappa}$ by two to improve the error by one order in
$1/N$ in the calculation of monomials of high order. This is the
reason for our not giving $w^u_3$ but only $w^u_4$ which yields
correct values for the integrals up to order $1/N^2$.

For COE and CSE, the situation is slightly more complicated. The
constraints on the matrices are not expressible in a simple way in
terms of products as done in Eqs.~(\ref{eq:cond}). It seems,
therefore, most convenient to limit the space of independent matrix
elements from the outset. In the case of the COE this is fairly
simple: The symmetry reduces the number of independent complex matrix
elements to $N(N+1)/2$. Therefore, we consider Gaussian averages in
the space of complex symmetric matrices $S = S^T$. We accordingly have
a contraction rule with two terms,
\begin{equation}
\int d\mu_g(S) S^*_{ij} S_{kl} = \frac{1}{N} \left (\delta_{ik}
\delta_{jl} + \delta_{il} \delta_{jk} \right) \ .
\label{eq:rule}
\end{equation}
The invariants are now defined in terms of $S S^{*}$. Introducing the
rule~(\ref{eq:rule}) into our program we can again calculate
$w^c_{\kappa}$. As an example we the coefficients $c^2_{\bf k}$ for
$w^c_2$,
\begin{eqnarray}
c^2_0 &=& 1 - \frac{N(N+1)}{4} \nonumber \\
c^2_1 &=& \frac{N+1}{2} \nonumber \\
c^2_2 &=& -\frac{(N+1)^3}{4N(N+3)} \nonumber \\
c^2_{1 1} &=& \frac{(N+1)^2}{4N(N+3)} \ .
\end{eqnarray}

In addition, the following subtle point must be considered. In
Section~\ref{wei} we have used the invariance of the Haar measure to
justify that contraction with $w_{\kappa}$ gives exact results for all
polynomials up to order $2 \kappa$ in the matrix elements. In the 
present case we have no invariance group and by consequence no Haar
measure. On the other hand Dyson's measure with respect to which we
integrate, is also totally defined by an invariance group albeit a
smaller one than that of ${\rm U}(N)$. The important point is that
again the measure is uniquely defined by a linear group of
transformations. The orthogonality conditions resulting from the
unitarity of the matrices are the same and symmetry is taken into
account explicitly in the contractions. Therefore all arguments again 
go through and indeed inspection of the results obtained by our code
with those obtained in Ref.~\cite{MS} shows agreement.

The case of CSE is simpler because it involves an invariance group,
but more complicated because the matrices are symplectic. Two ways
seem open to address this case. We might include the symplectic
property from the outset in the contraction rules, or we might
introduce this property as a constraint in the expression for the
weight function. While both ways seem possible it is not clear which
one is easier to follow. In view of the fact that CSE is of minor
importance for practical applications, we have left this problem open.
It is clear, however, that it can be tackled along the same lines.

\section{Conclusions}

We have presented a systematic way to calculate integrals over
monomials of matrix elements for compact matrix groups and for other
matrix ensembles whose measure is defined uniquely by an invariance
group, such as the circular orthogonal ensemble of unitary symmetric
matrices. This method gives exact results for monomials of low order.
For higher--order monomials, it leads to an error of order
$1/N^{\alpha}$ which is independent of the degree of the monomial. We
have given a lower bound on the integer $\alpha$, and we have shown
how $\alpha$ can be increased systematically. The method is
particularly suited for symbolic computer calculation. Codes are
available for ${\rm O}(N),\, {\rm U}(N)$ as well as for the circular
orthogonal ensemble in Mathematica and in C from one of the authors
(T.P.).

\section*{Acknowledgments} We wish  to thank B. Dietz and F. Leyvraz
for helpful discussions. T.H.S. wishes to thank the Humboldt 
Foundation for its generous support. T.P acknowledges financial support
by the Ministry of Education, Science and Sport of the 
Republic of Slovenia


\begin{thebibliography}{0}
\bibitem{ullah}  N.~Ullah and C.E.~Porter,  Phys. Rev. {\bf 132} 
(1963) 948.
\bibitem{MS} P.A.~Mello and T.H.~Seligman,
Nucl.Phys. A{\bf 344}, (1980) 498.
\bibitem{dyson} F.J.~Dyson, J. Math. Phys. {\bf 3} (1962) 120
\bibitem{gorin} T.~Gorin  (math-ph/0112012).
\bibitem{NT} T.~Guhr, A.~M\"uller-Groeling and H.A.~Weidenm\"uller,
Phys. Rep. {\bf 299}, 189 (1998).
\bibitem{PSW} T.~Prosen, T.H.~Seligman and H.A.~Weidenm\"uller
Europhys. Lett. {\bf 55} (2001) 12.
\end{thebibliography}
\end{document}